\documentstyle[12pt]{article}

\catcode`\@=11

\def\@normalsize{\@setsize\normalsize{15pt}\xiipt\@xiipt
\abovedisplayskip 14pt plus3pt minus3pt%
\belowdisplayskip \abovedisplayskip
\abovedisplayshortskip  \z@ plus3pt%
\belowdisplayshortskip  7pt plus3.5pt minus0pt}

\def\small{\@setsize\small{13.6pt}\xipt\@xipt
\abovedisplayskip 13pt plus3pt minus3pt%
\belowdisplayskip \abovedisplayskip
\abovedisplayshortskip  \z@ plus3pt%
\belowdisplayshortskip  7pt plus3.5pt minus0pt
\def\@listi{\parsep 4.5pt plus 2pt minus 1pt
            \itemsep \parsep
            \topsep 9pt plus 3pt minus 3pt}}

\def\underline#1{\relax\ifmmode\@@underline#1\else
	$\@@underline{\hbox{#1}}$\relax\fi}
\@twosidetrue

\relax

\catcode`@=12

\def\appendix{\setcounter{equation}{0}
               \setcounter{section}{0}
	\def\thesection{APPENDIX }
	\def\theequation{\Alph{section}.\arabic{equation}}}

\def\ps@headings{\def\@oddfoot{}\def\@evenfoot{}
\def\@oddhead{\hbox{}\hfill
	\makebox[.5\textwidth]{\raggedright\ignorespaces --\thepage{}--
	\hfill {}}}
\def\@oddhead{\hbox{}\hfill --\thepage{}-- \hfill
	{}}
\def\@evenhead{\@oddhead}
\def\subsectionmark##1{\markboth{##1}{}}
}

\ps@headings

\catcode`\@=12

\relax

\def\figcap{\section*{Figure Captions\markboth
	{FIGURECAPTIONS}{FIGURECAPTIONS}}\list
	{Fig. \arabic{enumi}:\hfill}{\settowidth\labelwidth{Fig. 999:}
	\leftmargin\labelwidth
	\advance\leftmargin\labelsep\usecounter{enumi}}}
 \relax
\def\tablecap{\section*{Table Captions\markboth
	{TABLECAPTIONS}{TABLECAPTIONS}}\list
	{Table \arabic{enumi}:\hfill}{\settowidth\labelwidth{Table 999:}
	\leftmargin\labelwidth
	\advance\leftmargin\labelsep\usecounter{enumi}}}
 \relax
\def\reflist{\section*{References\markboth
	{REFLIST}{REFLIST}}\list
	{[\arabic{enumi}]\hfill}{\settowidth\labelwidth{[999]}
	\leftmargin\labelwidth
	\advance\leftmargin\labelsep\usecounter{enumi}}}
 \relax

\catcode`\@=11

\def\ps@headings{\def\@oddfoot{}\def\@evenfoot{}
\def\@oddhead{\hbox{}\hfill
	\makebox[.5\textwidth]{\raggedright\ignorespaces --\thepage{}--
	\hfill {}}}
\def\@evenhead{\@oddhead}
\def\subsectionmark##1{\markboth{##1}{}}
}

\ps@headings

\relax

\newskip\humongous \humongous=0pt plus 1000pt minus 1000pt

\newif\ifdtup

\def\beq{\begin{equation}}
\def\eeq{\end{equation}}

\def\beqn{\begin{eqnarray}}
\def\eeqn{\end{eqnarray}}
\relax

\def\G2{{\; \rm GeV/}c^2}
\def\G{\; \rm GeV}

\def\dotx{\dotx{\dot\overline{x}}}

\relax

\begin{document}

\begin{titlepage}
\nopagebreak
\begin{flushright}

        {\normalsize     ITP--SB--91--40\\
                         March,~1992\\}

\end{flushright}

\vfill
\begin{center}
{\large \bf Anomalous Transformation  \\
                     in \\
            Supersymmetric
 Yang-Mills Theory}\footnote{Work supported in part by NSF Grant Phy 91-08054}

\vfill
        {\bf H.~Itoyama} \\
           and  \\
        {\bf B.~Razzaghe-Ashrafi}\\

        Institute for Theoretical Physics,\\
        State University of New York at Stony Brook,\\
        Stony Brook, NY 11794-3840, USA\\


\end{center}
\vfill

\begin{abstract}
 An  ``anomalous'' supersymmetry transformation
 of the gaugino axial  current   is given in supersymmetric Yang-Mills
theory.    The contact term is computed to one-loop
order by a gauge-invariant point-splitting procedure.
We reexamine the supercurrent anomaly  in this method.

\end{abstract}
\vfill

\end{titlepage}

 Much discussion has been given for
anomalies in supersymmetric field theories in various contexts
\cite{AGS,SA3/2,CPS,Konishi,GW,SUAN}.
Less attention has been paid, however, to quantum mechanical
 supersymmetry transformation laws which  composite operators obey
\cite{Konishi}.
  Short distance singularities demand
 that
these transformations  have an ``anomalous'' part-- more accurately--
the part which does not have a classical counterpart  and yet which is
necessary to preserve a proper quantum-mechanical transformation law.

In ref. \cite{Konishi},
Konishi has given an example in supersymmetric QCD. In this note,
we provide another example by examining the supersymmetry transformation
of the  gaugino axial current  in supersymmetric Yang-Mills theory.
We will see that our computation yields the following charge algebra
 (see later for the definitions):
\beqn
\label{eq:chargealgebra}
   [ {\bf Q}_{5}, {\bf Q}_{\alpha} ] =  {\bf Q}_{\alpha} - \int dx
 {\bf \Delta}_{0, \alpha}(x; \epsilon) \;\;\;.
\eeqn

We follow the notation of \cite{WB}.
We denote  operators in Heisenberg picture by bold faces whereas
operators in the interaction picture are denoted by  ordinary faces.
We will carry out computation by a gauge-invariant point-splitting procedure.

In order to find a connection  of our result with the existing literature
on the supercurrent/superconformal anomaly \cite{AGS,SA3/2,GW}, we have
 reevaluated the supercurrent anomaly
 by the same method.

\def\BL{\hbox{\boldmath $\lambda$}}

Let us begin with defining
 a regularized chiral  $U(1)$ current of gauginos $\BL,\bar{\BL}$
by a gauge-invariant point splitting:
\beqn
\label{eq:chiralreg}
{\bf j}_\ell (x \mid \epsilon) &\equiv& (\sigma_\ell)_{\alpha\dot{\alpha}} {\bf
j}^{\dot{\alpha}\alpha} (x \mid \epsilon),~~
 U (x \mid \epsilon) \equiv P \exp \left( - i g \int^{x+
\epsilon/2}_{x-\epsilon/2} dx'^m {\bf v}_m (x') \right), \nonumber \\
 \bar{\eta}_{\dot{\alpha}} {\bf j}^{\dot{\alpha}\alpha} (x \mid \epsilon)
\eta_\alpha &=& {1 \over k} tr \bar{\eta} \bar{\BL} (x+\epsilon/2)U
(x \mid \epsilon) \BL (x- \epsilon/2) \eta U^\dagger (x \mid \epsilon)\;,\;\;\;
  tr T^a T^b =  k\delta^{ab} \;\;\;.
\eeqn
The gauge fields ${\bf v}_{m}$ in the path-ordered exponentials are taken to
 be inert, namely, considered to appear only in external lines.
We should mention  that the current defined this way
has an anomalous divergence which  supposedly  obeys the Adler-Bardeen
theorem. This current is, therefore, quantum-mechanically
 distinct from the current
commonly known as $R$- current, which is  defined typically by
a (supersymmetric-)
 dimensional regularization, and  whose anomalous
divergence forms a supermultiplet with the superconformal anomaly
 and the trace anomaly \cite{GW}.

 A well-known supersymmetry algebra of elementary fields reads
\beqn
\label{eq:susy}
 [ \xi Q, {\bf v}^a_\ell (x)] = -i \xi \sigma_\ell \bar{\BL}^a (x)~,
{} ~&~&~ [ \bar{\xi} \bar{Q}, {\bf v}^a_\ell (x)]=
 i \bar{\xi} \bar{\sigma}_\ell \BL^a (x) \;\;\;, \nonumber \\
{}~ [ \xi Q, \BL^a (x) \eta ] = \xi \sigma^{nm} \eta {\bf v}^a_{nm} (x)~,
{}~&~&~ [ \bar{\xi} \bar{Q}, \bar{\BL}^a (x) \bar{\eta}] = \bar{\xi}
\bar{\sigma}^{nm} \bar{\eta}
{\bf v}^a_{nm} (x) \;\;\;,    \\
{}~ [ \bar{\xi} \bar{Q}, \BL^a (x) \eta ]~=~ [ \xi Q, \bar{\BL}^a (x)
\bar{\eta}]= &0& \;,\;\;  \nonumber
\eeqn
{}from which one finds
the classical transformation law for the axial current:
\beqn
\label{eq:clastrans}
 \delta_\xi {\bf j}_\ell = [\xi Q, {\bf j}_\ell] =-
\xi^\alpha {\bf S}_{\alpha \ell}\;\;,
{}~~{\rm with}~~~~
  \xi {\bf S}_\ell  \equiv - \xi \sigma^{nm} \sigma_\ell \BL^a {\bf v}^a_{nm}
\;\;\;.
\eeqn
 We will find its quantum-mechanical counterpart
shortly.

Let us introduce a
regularized supercurrent and its conjugate by
\beqn
\label{eq:supercurrent}
 \xi {\bf S}_\ell (x \mid \epsilon)  &\equiv&
- {1 \over k} tr \xi \sigma^{nm} \sigma_\ell
\bar{\BL}
 (x+ \epsilon/2) U (x \mid \epsilon) {\bf v}_{nm} (x- \epsilon/2) U^\dagger
(x \mid \epsilon)\;\;\;, \nonumber \\
 \bar{\xi} \bar{\bf S}_\ell (x\mid \epsilon) &\equiv& {1 \over k} tr \bar{\xi}
\bar{\sigma}^{nm} \bar{\sigma}_\ell \BL (x+ \epsilon/2) U
(x \mid \epsilon) {\bf v}_{nm} (x- \epsilon/2) U^\dagger (x \mid
\epsilon)\;\;\;,
\eeqn
We find a   supersymmetry algebra for the Heisenberg
operators:
\beqn
\delta_\xi {\bf j}_\ell (x \mid
\epsilon) &=& [\xi Q, {\bf j}_\ell (x \mid \epsilon)]
 =- \xi {\bf S}_\ell (x \mid \epsilon)+
\xi {\bf \Delta}_\ell (x \mid \epsilon) \;\;\;,
  \nonumber \\
 \delta_{\bar{\xi}} {\bf j}_\ell (x \mid \epsilon) &=& [\bar{\xi} \bar{Q}, {\bf
j}_\ell (x \mid \epsilon)]
 = \bar{\xi} \bar{\bf S}_\ell (x \mid \epsilon)-
 \bar{\xi} \bar{\bf \Delta}_\ell
(x \mid \epsilon)\;\;\;, \label{eq:Hsusyalg}
\eeqn
where
\beqn
\label{eq:Delta}
 \xi {\bf \Delta}_\ell
 (x \mid \epsilon)&=& {1 \over k} tr (\sigma_\ell \bar{\BL}
(x+ \epsilon/2))_\alpha \delta_\xi U(x;\epsilon) \BL (x-\epsilon/2)^\alpha
U^\dagger (x;\epsilon)  \;\;\; \nonumber \\
 &+&
 {1 \over k} tr (\sigma_\ell \bar{\BL} (x+ \epsilon/2))_\alpha U(x;\epsilon)
\BL (x-\epsilon/2)^\alpha \delta_\xi U^\dagger (x;\epsilon) \;\;\; \nonumber
  \\
{}~ &=& - ig \int^{x+\epsilon/2}_{x-\epsilon/2} dy_m {1 \over k} tr
(\sigma_\ell \bar{\BL}
 (x+\epsilon/2))_\alpha [\delta_\xi {\bf v}^m(y),
 \BL (x- \epsilon/2)^\alpha]   \nonumber  \\
&+&{\rm higher~order~expansion~in~} U
  \;\;\;.
\eeqn
We  call ${\bf \Delta}_{\ell}(x \mid \epsilon)$
  a contact term as it is cubic in fermions
and has a purely quantum-mechanical origin.
Defining
\beqn
\label{eq:Q5Q}
 {\bf Q}_5 \equiv \int d^{3}x {\bf j}_{0} (x \mid \epsilon)\;\;,\;
\;\; {\bf Q}_{\alpha} &\equiv& \int d^{3}x {\bf S}_{0 \alpha}
 (x \mid \epsilon)  =  Q_{\alpha} \;\;\;,
\eeqn
 we find the charge algebra stated in eq.~(\ref{eq:chargealgebra}).

Let us now evaluate this  term
 ${\bf \Delta}_{\ell} (x \mid \epsilon)$
 to lowest order in perturbation theory.
 We will use the  position space propagators for gauge bosons and gauginos:
\beqn
\label{eq:prop}
 <T \xi \lambda (x) \bar{\xi} \bar{\lambda}( y ) > &=& - \xi \sigma\cdot
 \partial  \bar{\xi} J_{1} \left( x-y \right)
  \;\;\;, \nonumber  \\
 < T v_{\ell}(x) v_{k}(y)> &=& -i[ \eta_{\ell k} J_{1}(x-y) + (1-\alpha)
 \partial_{\ell}^{(x)} \partial_{k}^{(x)}  J_{2} ( x-y) ]
\;\;\;, \nonumber \\
  J_{m} \left( x-y \right) &=& \int \frac{d^{4}k}{( 2\pi)^{4}}
 e^{-ik\cdot(x-y)} \frac{1}{(k^{2} - i0)^{m} } \;\;\;.
\eeqn
 The gauge parameter has been denoted by $\alpha$.
  The following formulas  are expedient to our calculation. ( Use, for
 instance, the symmetric
 integration method):
\beqn
\label{eq:contraction}
  \lim_{\epsilon \rightarrow 0} \epsilon^{\ell} \frac{\partial}
{ \partial \epsilon_{\ell^{\prime}} } J_{2}(\epsilon) &=& \frac{-i}{32\pi^{2}}
 \eta^{\ell \ell^{\prime}} \;\;\;, \nonumber \\
  \lim_{\epsilon \rightarrow 0} \epsilon^{\ell} \frac{\partial}
{ \partial \epsilon_{n} } \frac{\partial}
{ \partial \epsilon_{m} } \frac{\partial}
{ \partial \epsilon_{j} }  J_{3}(\epsilon) &=& \frac{i}{32\pi^{2}}
\frac{1}{6}
 ( \eta^{\ell n} \eta^{mj} + \eta^{\ell m} \eta^{ n j} + \eta^{\ell j}
\eta^{n m} )   \;\;\;.
\eeqn

In order to carry out, in the interaction picture,
 a normal-ordered (perturbative) expansion of the right
hand side of eq.~(\ref{eq:Delta})
 in the limit  $\epsilon_\ell \rightarrow 0$,  we need the following
 formula concerning the gaugino two-point
function:
\beqn
\label{eq:gauginotwo}
 \lim_{\epsilon = y-x  \rightarrow 0} \epsilon^\ell \bar{\eta} \bar{\BL}^a (y)
 \BL^b (x) \eta &=&
 -{ g \over 32 \pi^2} f^{abc} \partial_m v^c_n (x_0) \bar{\eta}
 K^{\ell mn}
(a',b') \eta + 0(g^3)  \;\;\;,  \\
 K^{\ell mn} (a',b') &=& \epsilon^{\ell m n r}
 \bar{\sigma}_r -i {(a'-b') \over 3}
(\eta^{\ell n} \bar{\sigma}^m + \eta^{\ell m} \bar{\sigma}^n + \eta^{m n}
\bar{\sigma}^\ell) \;\;\;, \nonumber
\eeqn
where
$ a'+b' = 1 $ and
$ x_0 = x+ b' \epsilon = y-a' \epsilon$.
Only the choice $a'=b'=1/2~,~ {\it i.e.}$~ the mid-point
prescription, leads to a gauge-covariant answer. We will adopt this choice in
what  follows.

We find
\beqn
\label{eq:answer}
 \lim_{\epsilon \rightarrow 0} \xi {\bf \Delta}_\ell (x \mid \epsilon)
 &=& {-ig^2 \over 32 \pi^2} C_2 (adj) \xi \sigma_j
\bar{\lambda}^c (x) \partial_m v^c_n tr \sigma_\ell K^{jmn}
 + 0 (g^4) \;\;\; \nonumber \\
{}~&=& {-i g^2 \over 16
\pi^2} C_2 (adj) \xi \sigma^{j} \bar{\lambda}^c (x)
 \tilde{v}^{c}_{\ell j} (x)
 + 0(g^4) \;\;\;,
\eeqn
 where
\beqn
\label{eq:dualdef}
\tilde{v}_{\ell j}^{c} \equiv \frac{1}{2} \epsilon_{\ell jmn}v^{mn~c} \;\;\;.
\eeqn

Our result eq.~(\ref{eq:answer}) is easily seen to be proportional to
the supersymmetry  transformation of the Chern-Simons density, which
 are (in our normalization)
\beqn
\label{eq:chernsimons}
 K^{\ell} (x) &\equiv& \frac{g^{2}}{ 4 \pi^{2}}  \epsilon^{\ell
mnr}  \frac{1}{k} tr\left( v_{m}(x) \partial_{n}v_{r}(x) + \frac{2}{3}
ig v_{m}v_{n}v_{r} \right) \;\;\;, \nonumber \\
 \delta_{\xi} K_{\ell}(x) &=& \frac{-ig^{2}} {2 \pi^{2}}
 \xi \sigma^{j} {\bar \lambda}^{a} (x) \tilde{v}_{\ell j}^{a} (x)\;\;\;.
\eeqn
 The coefficient in eq.~(\ref{eq:answer})
 differs, however, from  what one would
 naively infer from the axial anomaly:
\beqn
\label{eq:axialan}
  \lim_{\epsilon \rightarrow 0}
\partial^{\ell} {\bf j}_{\ell}(x\mid \epsilon)
=   \frac {1}{2}  C_{2}(adj) \partial^{\ell} K_{\ell} (x) \;\;\;.
\eeqn
( The extra factor $1/2$ is accounted for by Majorana fermions.)
 This is  as it should be : taking derivatives increases  degrees of
ultraviolet divergences in momentum integrands and does not commute with
 the limit $\epsilon \rightarrow 0$.

In order to understand our result and its connection to the existing
literature better, we  will calculate  the divergence of the supercurrent
 ${\displaystyle \lim_{\epsilon \rightarrow 0} }
 \partial^{\ell} \xi \cdot {\bf S}_{\ell} (x \mid \epsilon)$
    by the present method. ( ${\bar \xi} {\bar \sigma} \cdot {\bf S} (x
\mid \epsilon) =0 $ in the present regularization.)
 After using equations of motion and Bianchi identity, \footnote{ As we have
 used the equation of motion for the gauge fields without a gauge fixing term,
 eq.~(\ref{eq:superdiv}) should be viewed as a statement with respect to
 the physical states.  }  we find
\beqn
\label{eq:superdiv}
   \partial^{\ell} \xi \cdot {\bf S}_{\ell}
(x \mid \epsilon)
{}~&=&  \frac{ig}{k} tr \xi \sigma^{mn} \sigma^{\ell}
{\bar {\BL}} (x + \epsilon/2)
  \left[ \partial_{\ell} \left( \int^{x+\epsilon/2}_{x- \epsilon/2}
 dx^{\prime}_{j} {\bf v}^{j} (x^{\prime}) \right), {\bf v}_{mn}
 (x-\epsilon/2) \right]
 \;\;\; \nonumber \\
{} ~&+&~{\rm higher~ order~ expansion~in~}~U \;\;\;.
\eeqn
 Going to the interaction picture, we find, up to order $ g^{2}$,
\beqn
\label{eq:superdiv2}
  \lim_{\epsilon \rightarrow 0}
 \partial^{\ell} \xi \cdot {\bf S}_{\ell} (x \mid \epsilon)
= -g^{2} C_{2}(adj) \lim_{\epsilon \rightarrow 0}
 \epsilon^{j} \partial_{\ell} v_{j}^{a} (x- \epsilon/2) \times  \nonumber \\
\int d^{4}z ( \sigma^{k} {\bar \lambda}^{a}(z) )_{\alpha}
  <T \xi \sigma^{mn} \sigma^{\ell}  {\bar \lambda}(x+\epsilon/2)
 \lambda^{\alpha} (z)>     < T v_{mn} (x-\epsilon/2) v_{k} (z)> \;\;\;.
\eeqn
  Converting the expression
 into  momentum space,
we find
\beqn
\label{eq:suanswer}
  \lim_{\epsilon \rightarrow 0}
 \partial^{\ell} \xi \cdot {\bf S}_{\ell} (x \mid \epsilon)
 &=&    \frac{g^{2}}{32\pi^{2}} C_{2}(adj)
 \xi  (\partial^{\ell} \sigma \cdot v^{a} (x))
 {\bar \sigma} \cdot \partial \sigma_{\ell}
{\bar \lambda}^{a} (x)  \;\;\; \nonumber \\
 &=& \frac{g^{2}}{32\pi^{2}} C_{2}(adj)
[ (\partial \cdot v^{a}(x)) ( \xi \sigma \cdot \partial {\bar \lambda}^{a}
(x) )  \nonumber \\
{}~&~ -& (\partial_{ \{ \ell } v_{ m \}} (x) )( \xi \sigma^{\ell} \partial^{m}
{\bar \lambda}^{a} (x) ) -i \tilde{v}_{\ell m}^{a} ( \xi \sigma^{\ell}
\partial^{m} {\bar \lambda}^{a} (x) ) ] \;\;\;,
\eeqn
 where we used  eqs.~(\ref{eq:prop}),(\ref{eq:contraction})and the midpoint
prescription.
 The right hand side of
 eq.~(\ref{eq:suanswer}) can be written
 as a total derivative
  $\partial^{\ell} \left( \frac{g^{2}}{32\pi^{2}} C_{2}(adj)
 \xi   \sigma \cdot v^{a} (x)
 {\bar \sigma} \cdot \partial \sigma^{\ell}
{\bar \lambda}^{a} (x) \right) $ once the on-shell condition for gauginos
 $ \sigma \cdot \partial {\bar \lambda}^{a} (x) =0$ is imposed.

Separately, we find
\beqn
\label{eq:deltadiv}
 \lim_{\epsilon \rightarrow 0} \left( \xi \partial^{\ell} {\bf \Delta}_{\ell}
(x \mid \epsilon) \right) = \frac{ig^{2}}{16 \pi^{2}} C_{2}(adj)
\left( \xi \sigma^{j} \partial^{\ell} {\bar \lambda}^{a} (x) \right) \tilde
{v}_{j \ell}^{a} (x) \;\;\;.
\eeqn
Therefore,
\beqn
\label{eq:finalsuper}
 \lim_{\epsilon \rightarrow 0} \partial^{\ell} \delta_{\xi}
 {\bf j}_\ell (x \mid \epsilon) &=&
 \lim_{\epsilon \rightarrow 0} \partial^{\ell} \left( -\xi \cdot {\bf S}_{\ell}
(x \mid \epsilon)  +  \xi  {\bf \Delta}_{\ell}
(x \mid \epsilon)  \right)  \nonumber \\
 &=&    \frac{g^{2}}{32\pi^{2}} C_{2}(adj) c_{1}
[ - (\partial \cdot v^{a}(x)) ( \xi \sigma \cdot \partial {\bar \lambda}^{a}
(x) )  \nonumber \\
 &+& (\partial_{ \{ \ell }  v_{ m \} }^{a} (x) )
 ( \xi \sigma^{\ell} \partial^{m}
{\bar \lambda}^{a} (x) ) - 3i \tilde{v}_{\ell m}^{a} ( \xi \sigma^{m}
\partial^{\ell} {\bar \lambda}^{a} (x) ) ] \;\;\; \\
 &\neq&
 \delta_{\xi}  \lim_{\epsilon \rightarrow 0}  \partial^{\ell}
 {\bf j}_\ell (x \mid \epsilon)  \;\;\;. \nonumber
\eeqn
  We  see that the
supersymmetry  transformation does not commute with the limiting procedure.
 The first term in eq.~(\ref{eq:finalsuper}) vanishes on-shell whereas
the second term can be written as a total derivative on-shell and therefore
  can formally be absorbed into a redefinition of
$\xi \cdot {\bf S}_{\ell} (x \mid \epsilon)$.
The  supercurrent anomaly  on-shell   can therefore be written  as
\beqn
\label{eq:onshell}
     \frac{-3ig^{2}}{32\pi^{2}} C_{2}(adj)
 \partial^{\ell} (  \tilde{v}_{\ell m}^{a} \xi \sigma^{m}
 {\bar \lambda}^{a} (x) ) \;\;\;.
\eeqn
 This agrees with  eq.~(4.11) of the first reference of~\cite{AGS}
 conjectured to be equivalent to
\beqn
\label{eq:another}
     \frac{3g^{2}}{32\pi^{2}} C_{2}(adj)
 \partial^{\ell} (  v_{\ell m}^{a} \xi \sigma^{m}
 {\bar \lambda}^{a} (x) ) \;\;\;.
\eeqn
It is interesting to see that the contact term is essential
in saturating  the requisite
coefficient of the anomaly\footnote{ Eq.~(\ref{eq:another})
 was derived from
 the calculation of the on-shell matrix elements of
the supercurrent divergence
 by the Adler-Rosenberg method and the supersymmetric dimensional
regularization \cite{AGS,SA3/2}. }.

To summarize, a current defined by
\beqn
\label{eq:finaldef}
\xi  \hat{ {\bf S}}_{\ell}(x \mid \epsilon)
 &\equiv&  \delta_{\xi} {\bf j}_\ell (x \mid \epsilon)
  - \frac{g^{2}}{32\pi^{2}} C_{2}(adj)
[ \xi \sigma_{\ell} v^{a}(x) \cdot \partial {\bar \lambda}^{a}(x)
 + \xi \sigma \cdot  v^{a}(x)  \partial_{\ell} {\bar \lambda}^{a}(x) ]
  \nonumber \\
    &+& \frac{3i g^{2}}{32\pi^{2}} C_{2}(adj)
 \tilde{v}_{\ell m}^{a}  \xi \sigma^{m} {\bar \lambda}^{a} (x) \;\;\;.
\eeqn
is conserved and saturates the superconformal anomaly:
\beqn
\label{eq:sanomaly}
{\bar \xi} {\bar \sigma} \cdot \hat{ {\bf S}}(x \mid \epsilon)
=    - \frac{3g^{2}}{16\pi^{2}} C_{2}(adj)
    v_{\ell m}^{a} {\bar \xi} {\bar \sigma}^{ \ell m}
 {\bar \lambda}^{a} (x)  \;\;\;.
\eeqn

We have not fully
 investigated implications of eq.~(\ref{eq:chargealgebra}).
One obvious thing is, however,  that the supercharge $Q_{\alpha}$
does not carry a definite chiral charge. For the vacuum with unbroken
supersymmetry,
\beqn
\label{eq:conseq}
  \delta_{\xi} {\bf Q}_{5} \mid 0 > = - \int dx \xi {\bf \Delta} (x) \mid
  0 > \;\;\;
\eeqn
 holds.
  This illustrates a point difficult to incorporate in the Born-Oppenheimer
 approximation \cite{BO}. (See also \cite{W,CD}.)
A mere truncation to the zero momentum modes does not respect this equation.
 Additional insertions into the ground state wave function are required
in order to implement eq.~(\ref{eq:conseq}).

We thank Marc Grisaru for a useful discussion.

\end{document}